\documentclass[aps,prl,twocolumn,superscriptaddress,preprintnumbers,nofootinbib]{revtex4-1}
\usepackage{graphicx,amssymb,amsmath,color,slashed}
\usepackage{enumerate}
\usepackage{hyperref}

\def\beq{\begin{equation}}
\def\eeq{\end{equation}}

\def\bea{\begin{eqnarray}}
\def\eea{\end{eqnarray}}
\def\bei{\begin{itemize}}
\def\eei{\end{itemize}}
\def\bmat{\begin{matrix}}
\def\emat{\end{matrix}}
\def\ble{\begin{flushleft}}
\def\ele{\end{flushleft}}
\def\={\,=\,}
\def\+{\,+\,}
\def\-{\,-\,}

\def\GeV{\,{\rm GeV}\,}
\def\TeV{\,{\rm TeV}\,}
\def\fb{\, {\rm fb} \,}

\def\MET{E_T^{\textrm{miss}} }

\def\xfb{{\rm fb}}

\newcommand{\Fig}[1]{Fig.~\ref{#1}}
\newcommand{\Eq}[1]{Eq.(\ref{#1})}

\begin{document}

\title{Gaugino physics of split supersymmetry spectrum \\ at the LHC and future proton colliders}

\author{Sunghoon Jung}
\affiliation{School of Physics, Korea Institute for Adavanced Study, Seoul 130-722, Korea}

\author{James D. Wells}
\affiliation{Physics Department, University of Michigan, Ann Arbor, MI USA}

\begin{abstract}
Discovery of the Higgs boson and lack of discovery of superpartners in the first run at LHC are both predictions of split supersymmetry with thermal dark matter. We discuss what it would take to find gluinos at hadron supercolliders, including the LHC at 14 TeV center of mass energy, and future $pp$ colliders at 100 TeV and 200 TeV. We generalize the discussion by re-expressing the search capacity in terms of gluino to lightest superpartner mass ratio, and apply results to other scenarios, such as gauge mediation and mirage mediation.
\end{abstract}

\preprint{KIAS-P13064}

\maketitle

%%%%%%%%
\section{Introduction}

The split SUSY spectrum of lighter fermions and heavier scalars is an attractive possibility~\cite{Wells:2003tf,ArkaniHamed:2004fb,Giudice:2004tc,Wells:2004di}. Phenomenologically, heavy scalars help to avoid bounds from LHC searches, flavor physics constraints from Kaon and B mesons, and limits from electric dipole moments of electrons and neutrons. The 125 GeV Higgs boson mass is accommodated easily by heavy scalars although some degree of fine tuning between mass parameters exists in triggering electroweak symmetry breaking (EWSB). 

Importantly, split scenarios are not arbitrarily heavy when reasonable assumptions are applied. Too heavy scalars would induce too much quantum corrections to the Higgs boson mass. 100 ($10^5$) TeV stops can induce 125 GeV Higgs mass for $\tan \beta \sim 5 (2)$ \cite{Giudice:2011cg,Ibe:2011aa}, for example. Also, too heavy LSPs would have too much relic abundance by now. Including Sommerfeld enhancement, 3.1 (1.0) TeV wino (higgsino) LSP can have the right amount of relic density~\cite{Cohen:2013ama,Fan:2013faa}. A heavier LSP would be generally constrained -- we are ignoring  possibilities of conspired mechanisms to dilute relic density. Gauge coupling unification prefers gauginos and higgsinos to be below 5-50 TeV~\cite{Giudice:2004tc,Arvanitaki:2012ps}.

The general existence of upper bounds on the split spectrum, especially the LSP in the TeV region, may imply that we can test the split scenario as a whole in near-future collider experiments. Scalars of ${\cal O}(100-1000)\TeV$ are too heavy for collider discovery, although they may show evidence in future flavor and CP violation measurements for some cases. Indirect detections of heavy wino dark matter are other useful probes~\cite{Cohen:2013ama,Fan:2013faa}, but are subject to various uncertainties, some maybe even unknown, intrinsic to astrophysical observations. Direct probes are still wanted, and lighter gauginos can play an important role in this endeavor.

Given these, what do we really need to eventually probe split scenarios up to the relic density upper bounds on LSP masses at a future collider?

Scenarios leading to split spectrum typically generate gaugino masses from anomaly mediation (AMSB)~\cite{Randall:1998uk,Giudice:1998xp}. It is because if SUSY is spontaneously broken by charged superfields, for example, gaugino masses are prohibited while scalar masses are not. A loop-factor hierarchy between fermions and scalars is generated. {Some variants are discussed in Refs.\cite{Ibe:2006de,ArkaniHamed:2012gw,Arvanitaki:2012ps}.}

We start by discussing the AMSB gaugino model without light higgsinos. This model is not only generic, but provides a simple and clear framework for our analysis whose results become the basis for more general studies. Most simplifying assumptions that we will discuss for collider and cosmological bounds are automatically satisfied in this model. This model may also be the most difficult scenario for the discovery as the gluino takes on its heaviest value with respect to the wino mass, and the winos are degenerate and difficult to find at colliders directly. Thus, studying this scenario can reasonably provide farthest discovery reach estimation. 

After this discussion, we generalize the spectrum and investigate how light higgsino NLSPs added to the AMSB gaugino spectrum modify the analysis, and how we can apply the results to a general spectrum with variable mass ratios and ordering so that other SUSY breaking mediation models can be studied. Model-dependent cosmological bounds are not considered together. We demonstrate that a useful parameter is gluino-to-LSP mass ratio both for the discovery and the discrimination of models.

%%%%%%%%%%%%%%%%%%%%%%%%%%%%%%%%%%%%%%%
\section{AMSB gauginos without higgsinos}

\begin{figure*}
\centerline{
\includegraphics[width=0.98\columnwidth]{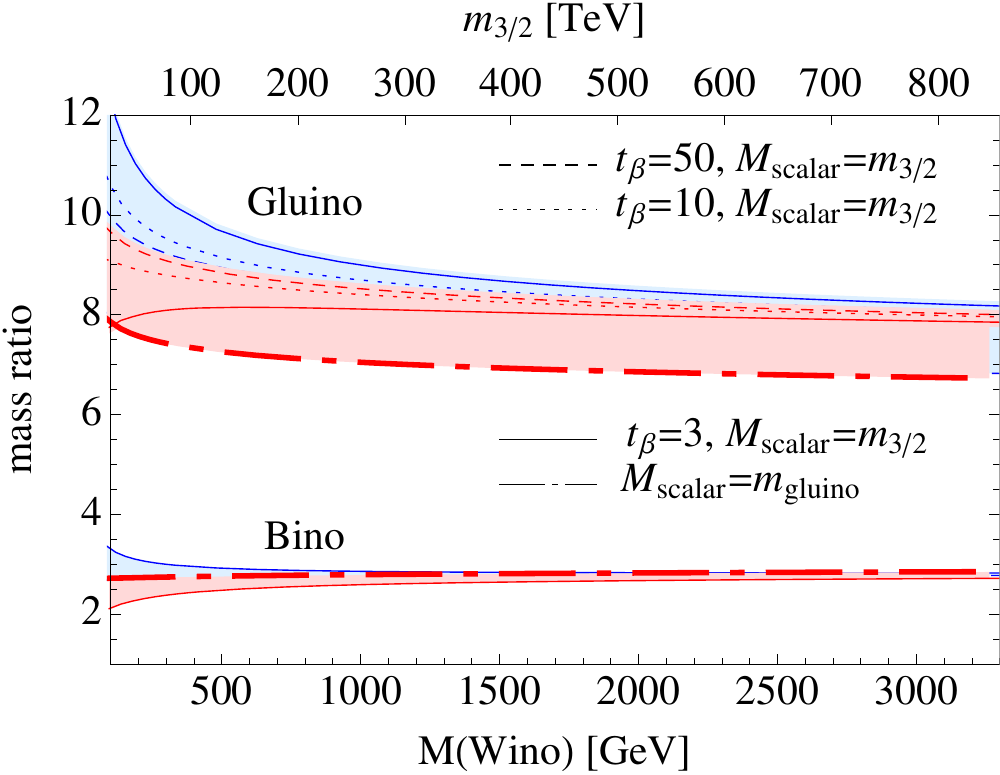}
\includegraphics[width=0.98\columnwidth]{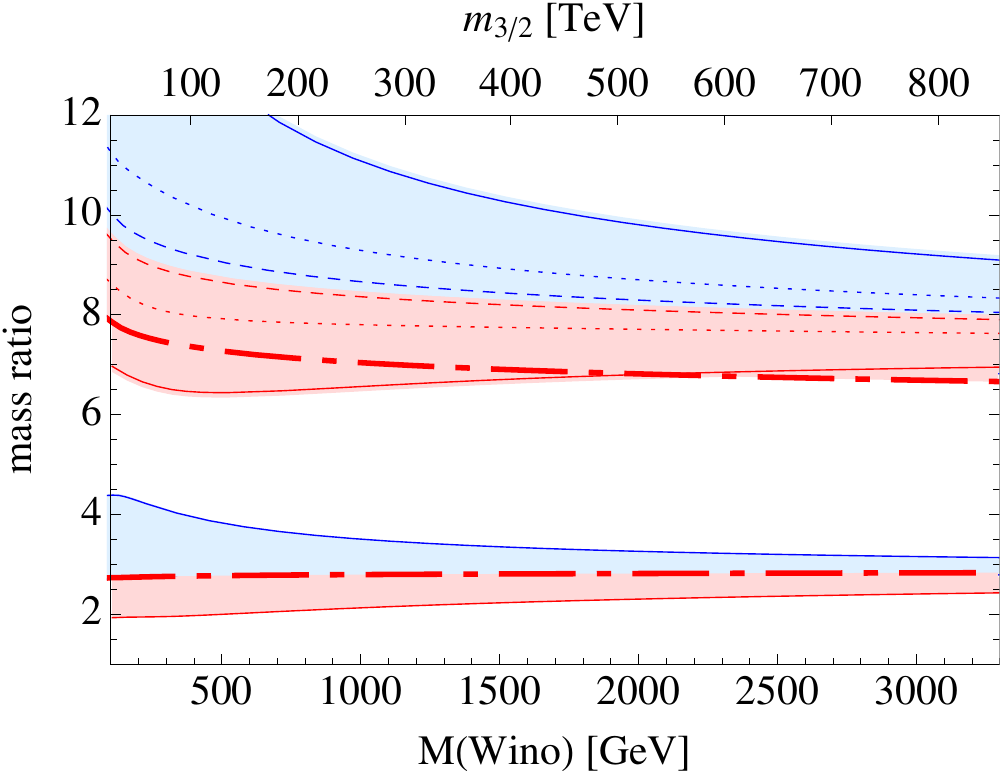}
}
\caption{AMSB gaugino mass ratios at NLO with higgsino $|\mu|=$4 TeV {\bf (left)} and 50 TeV {\bf (right)}. Red (blue) lines are with positive (negative) $\mu$. Colored bands are spanned by varying $\tan \beta$ ($3 \leq \tan \beta \leq 50$), scalar masses ($m_{\widetilde{g}} \leq m_{\widetilde{f}} \leq m_{3/2}$) and renormalization scale ($Q_0/2 \leq Q \leq 2Q_0$). Thick dot-dashed line is with light scalar masses $m_{\widetilde{f}}=m_{\widetilde{g}}$, and all other lines are with heavy scalar masses $m_{\widetilde{f}}=m_{3/2}$. Only a few lines are shown for bino for better readability. In the upper horizontal axis, we also show corresponding $m_{3/2}$ for $\tan \beta=50$ and $m_{\widetilde{f}}=m_{3/2}$.}
\label{fig:gaugino_masses}
\end{figure*}
%

%%%%%
\vspace{0.1in}
{\bf \emph{AMSB spectrum.}}
In a simple model we consider in this section, scalars and higgsinos are all heavy, and lighter AMSB gaugino masses at leading order (LO) are
\beq
M_i^A \= \frac{ b_i \alpha_i}{4\pi} m_{3/2}.
\label{eq:amsb} \eeq
$b_i$ are beta function coefficients. Wino is the LSP and gluino is heaviest. Next-to-leading order (NLO) results are used in this paper and are plotted in \Fig{fig:gaugino_masses}; see appendix for how we obtain them. Notably, NLO mass ratios depend somewhat on scalar masses, higgsino masses, and the overall wino mass scale which can be important for split spectrum. With heavy scalars, higgsino mass dependence is largest for small $\tan \beta$ -- higgsino threshold correction, $\delta M_{1,2} \sim -\frac{\alpha_{1,2}}{8\pi} \, 2\mu \sin 2\beta \, \log \frac{\mu^2}{m_0^2}$, proportional to $\mu$ rather than $M_{1,2}$ can be thought of as renormalization group-induced mixing effect~\cite{Giudice:2004tc}. When $\tan \beta$ is large, heavy squark threshold corrections raise mainly only the gluino mass, thus raising the mass ratio. If scalars are light, effects from higgsinos and scalars are small -- the simplest AMSB models predict scalars as heavy as the gravitino~\cite{Giudice:1998xp,Wells:2004di,Ibe:2011aa}, $m_{3/2}$, but making them lighter is also discussed in light of fine tuning issue~\cite{Randall:1998uk}. Higgsinos heavier than about 50 TeV may be disfavored by gauge coupling unification and  are not considered; heavier higgsinos could have distorted the spectrum more~\cite{ArkaniHamed:2012gw,Ibe:2006de}. When $\mu$ is negative, shown as the blue region in the figures, the wino gets negative corrections, raising the gluino to wino mass ratio further.

The uncertainty colored bands are spanned by mentioned variations of scalar masses ($m_{\widetilde{g}} \leq m_{\widetilde{f}} \leq m_{3/2}$), $\tan \beta$ ($3 \leq \tan \beta \leq 50$) as well as renormalization scale ($Q_0/2 \leq Q \leq 2Q_0$). The central scale is chosen $Q_0 = M_i^{LO}(Q_0)$ to minimize a log. The uncertainty from the renormalization scale is smallest among these. The rapid change in the light wino region is due to gauge coupling running (with $Q$) mixed with the aforementioned effects.

In this section, we assume heavy $m_{\widetilde{f}} = m_{3/2}$ and large $\tan \beta =50$. Later, we discuss how uncertainties from model parameters can be reflected in the disovery prospect.

Heavy higgsinos can still help to achieve radiative EWSB. If $m_{H_u}^2$ runs quickly down to an approximate infrared fixed point with value $-m_{\widetilde{t}_{L,R}}^2<0$, a similar size of $|\mu|^2 \sim |m_{H_u}^2|$ can trigger EWSB (at least for large $\tan \beta$)
\beq
m_Z^2 \, \sim \,  -2 (m_{H_u}^2 + |\mu|^2) + 0.5 \sin^2 (2\beta)  (m_{H_d}^2 - m_{H_u}^2 )
\eeq
leaving a light Higgs boson
\beq
\det \left( \bmat |\mu|^2 + m_{H_u}^2 & -B_\mu \\ -B_\mu & |\mu|^2 + m_{H_d}^2 \emat \right) \simeq 0.
\eeq

%%%%%%%%%%%%%%%%%%%%%%%%%%%%%%%%%%%%%%%%%%%%%%
\vspace{0.1in}
{\bf \emph{Gluino pair with effective mass.}}
Gauginos are well-separated in mass, and rarely mix. The LSP is thus nearly a degenerate wino. Wino production does not generate visible hard objects. Drell-Yan bino production is suppressed as the bino does not couple to $W,Z$ bosons. In this scenario they are not suitable production modes for discovery at a hadron collider.

Gluino pair production may be the only viable discovery channel at a hadron collider. To begin with, the gluino is assumed to decay as
\beq
\widetilde{g} \to \widetilde{w} \, q \bar{q},  \qquad q = u, d, s, c, b, t
\label{eq:gluinodecay}\eeq
via off-shell squarks. It leads to traditional SUSY signatures of jets plus missing energy. We do not distinguish chargino and neutralino of the same kind because nearly degenerate chargino decays produce only invisibly soft particles.

Decays into top quarks (either into $t\bar{t} \chi_1^0$ or $t \bar{b} \chi_1^-$) are not only possible but can be enhanced if stops are lighter than squarks. However, top quarks are highly boosted in heavy gluino decays and top decay products are collimated within an opening angle $\Delta R \sim m_t/p_T(t) \sim 3m_t / M_{\widetilde{g}} \ll 0.4$ -- where 0.4 is our jet radius in jet algorithm. Thus, without a dedicated top tagging technique applied, top would look essentially like a single jet. We do not distinguish decays into top quarks from other light quarks in this paper. Likewise, we do not apply $b$-tagging.

Only a small fraction of gluinos decay into intermediate binos. Loop-induced two-body decay, $\widetilde{g} \to g \widetilde{w}$, is small when gluino is multi-TeV~\cite{Toharia:2005gm,Gambino:2005eh}. We ignore these decay modes in this paper.

In all, the assumption of \Eq{eq:gluinodecay} is reasonable.

It has been widely studied that high-mass supersymmetry can be efficiently probed by variables sensitive to the heaviness and hardness of SUSY particles. One typical example variable is effective mass, $M_{eff}$~\cite{Hinchliffe:1996iu}:
\beq
M_{eff} \= \sum_i p_T(i) \+ \MET,
\eeq
where the scalar sum runs over all jets with $p_T>50\GeV,\, \eta<5.0$ and leptons with $p_T>15\GeV,\, \eta<2.5$. Main backgrounds to this analysis are $W+j$, $Z+j$ and top pair. All signal and background event samples are generated using {\sc MadGraph}~\cite{Alwall:2011uj} and Pythia~\cite{Sjostrand:2006za}. We refer to the Appendix on how we reconstruct physical particles and generate event samples.

Gluino pair production rate is multiplied by a constant K-factor 2 -- no result is available yet, but 2 seems reasonable from Ref.\cite{Beenakker:1996ch}. Background rates are normalized to Pythia-matched cross-sections that approximately take into account some of the NLO corrections.

\begin{figure}
\centerline{
\includegraphics[width=0.98\columnwidth]{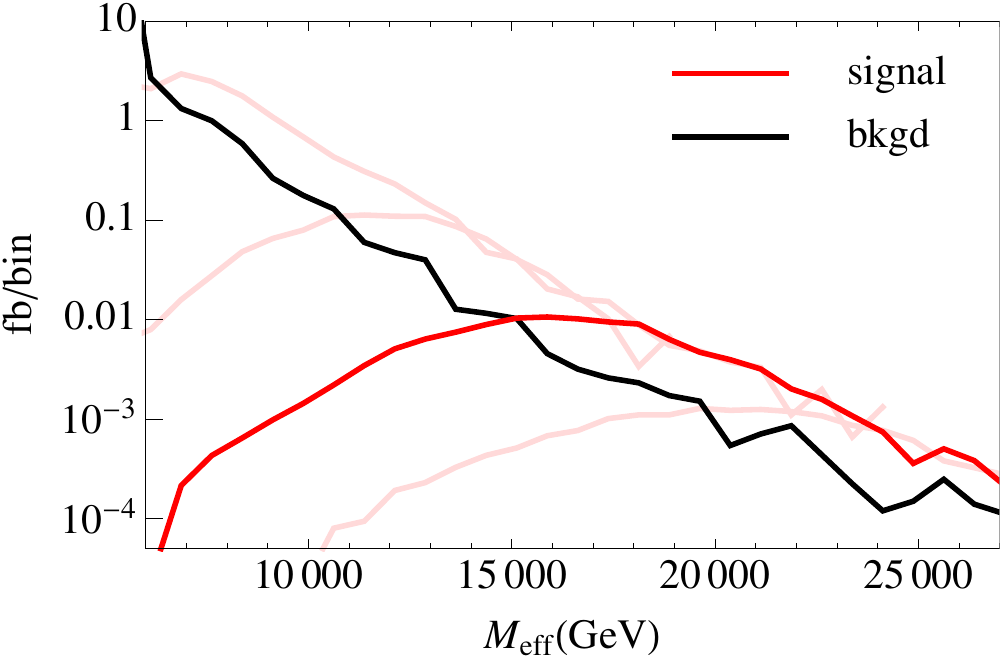}
}
\caption{Effective mass distribution after our discovery cuts except for the $M_{eff}$ cut. Several AMSB signals are shown together: dark red from 10 TeV gluino pair, and other light reds from 5, 7.5, 12.5 TeV gluino pair. Background (black) includes $Z+j$, top pair and $W+j$. Gluino decays as $\widetilde{g} \to q \bar{q} \chi_1$. 100 TeV $pp$ collider is assumed.}
\label{fig:meff}
\end{figure}

Based on baseline cuts developed by Hinchlifffe and Paige~\cite{Hinchliffe:2001bz}, we devise and optimize discovery cuts for 10 TeV gluino at 100 TeV $pp$ collision:
\bei
\item At least two jets with $p_T > 0.1\, M_{eff}$
\item lepton veto : definition of an isolated lepton is described in appendix.
\item $\MET > 0.2 M_{eff}$ and $p_T(j_1) < 0.35\, M_{eff}$
\item $\Delta \phi(j_1,\MET) < \pi -0.2$ and $\Delta \phi(j_1,j_2)<2\pi/3$
\item $M_{eff}> c M_{\widetilde{g}}$ where a constant $c$ is optimized.
\eei
After first four set of cuts are applied, we optimize the last cut on $M_{eff}$ to maximize statistical significance while requiring at least 10 signal events. See \Fig{fig:meff} for the spectrum after first four cuts. The optimal constant $c=1.5$ is found. We obtain $\sigma_S \simeq 0.072\fb$ and $\sigma_B \simeq 0.025\fb$ generating 11 signal events and a statistical significance $\sim 5.5$ with 150 fb$^{-1}$. $S/B \simeq 3$ is well above systematic uncertainties. Dominant backgrounds after all cuts are $Z+j$ and $t\bar{t}$ subdominantly -- $W+j$ is suppressed by the lepton veto.

We have also considered more sophisticated variables such as jet mass, $M_J = \sum_{i \in J} m(j_1)$, and (sub-)jet counting~\cite{Hook:2012fd,Hedri:2013pvl}. However, they do not work better than the simple $M_{eff}$ in our case. These variables were designed to work best for a high-multiplicity environment, such as for 10-jet final states. But our gluino pair produces only 4 quarks at leading order, and these other techniques are not optimal for our low multiplicity signal.

\begin{table}[t] \centering
\begin{tabular}{c|c|c|c|c|c}
\hline \hline
$\sqrt{s}$ & $m_{\widetilde{g}}$ & $\sigma_S$ & ${\cal L}$ & $S/B$ & stat. sig. \\
\hline \hline
100 TeV & 5 TeV & 21 fb (LO) & 5/fb & 2.9 & 12 \\
100 TeV & 10 TeV & 0.15 fb (LO) & 150/fb & 3.0 & 5.5 (5.6)\\ 
200 TeV & 15 TeV & 0.27 fb (LO) & 120/fb & 3.2 & 7 (6.7)\\
\hline \hline
\end{tabular}
\caption{Comparison of numerically optimized significance (last column without paranthesis) and the ones obtained from the scaling rule applied to the results in the first row (this estimation is shown in paranthesis in the last column).}
\label{tab:scaling}\end{table}
%
%%%%%%%%%%%%%%%%%%%%%%%%%%%%%%%%
\vspace{0.1in}
{\bf \emph{Scaling rule and discovery reach.}}
Remarkably, the sensitivity to gluino mass scales in a simple way with gluino mass and collision energy. \Fig{fig:meff} shows that signal-to-background ratios to the right of the signal peak are almost constant after discovery cuts for each gluino mass. This can be understood as gluinos becoming effectively massless at highest values of $M_{eff}$. Cut efficiencies should also be constant.

If these hold true, a simple scaling rule of the statistical significance can be obtained. The statistical significance of the signal with gluino mass $m_i$, production rate $\sigma_{Si}$ and luminosity ${\cal L}_i$ is
\bea
({\rm significance})_i & = & \frac{\sigma_{Si}\, \epsilon_{Si}}{\sqrt{\sigma_{Bi} \, \epsilon_{Bi}}} \sqrt{ {\cal L}_i}
\nonumber \\ &= & \sqrt{A_i \epsilon_{Si}} \sqrt{\sigma_{Si} {\cal L}_i }
\eea
where $\sigma_{B_i}$ is total background rate, $\epsilon_i$ is the cut efficiency, and $A\equiv  \sigma_S \epsilon_S / \sigma_B \epsilon_B$.  Since $A$ and $\epsilon_{Si}$ stay constant over the interesting range of gluino mass, the statistical significance scales as
\beq
\frac{ ({\rm significance})_i}{({\rm significance)}_j} \= \sqrt{ \frac{ \sigma_{Si} {\cal L}_i }{ \sigma_{Sj} {\cal L}_j } }.
\label{eq:scaling} \eeq 
We checked that this scaling rule approximately applies to a wide range of collision energies ($\sqrt{s} = 40 \sim 200$ TeV at least) and gluino masses ($\sqrt{s} = 3 \sim 20$ TeV at least) with split SUSY spectrum; see Table~\ref{tab:scaling}.

Although it is approximately valid, the scaling rule is very convenient because once sensitivity is estimated for one gluino mass at a certain energy and luminosity, the scaling can be used to predict sensitivities to a wide ranges of masses, energies and luminosities without needing to repeat lengthy numerical analyses. 

\begin{figure}
\centerline{
\includegraphics[width=0.98\columnwidth]{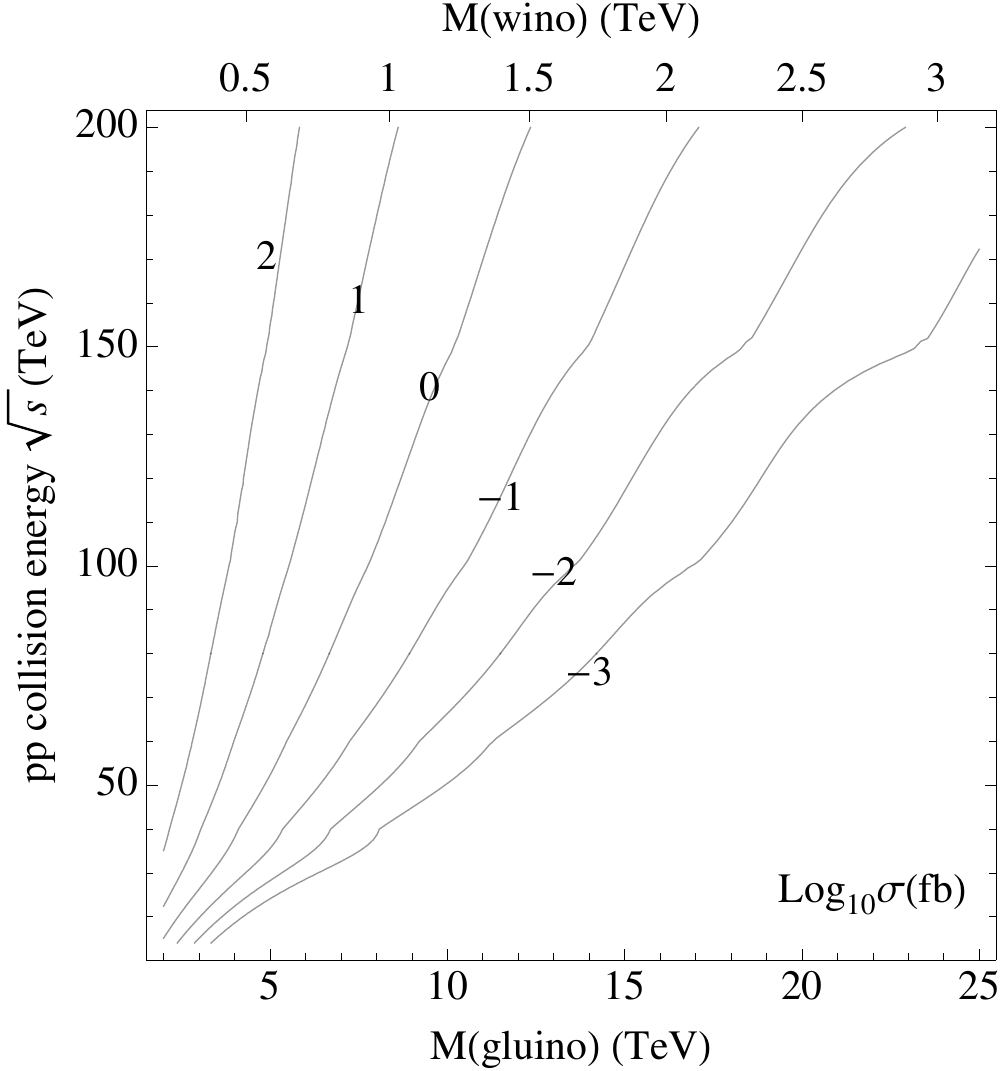} 
}
\caption{Gluino pair production rate (fb) at $pp$ collision. Leading order cross-sections. In the upper horizontal axis, the corresponding wino mass is also shown in the AMSB model with $m_{\widetilde{f}}=m_{3/2}$ and $\tan \beta =50$. }
\label{fig:gluino-prod}
\end{figure}
\begin{figure}
\centerline{
\includegraphics[width=0.98\columnwidth]{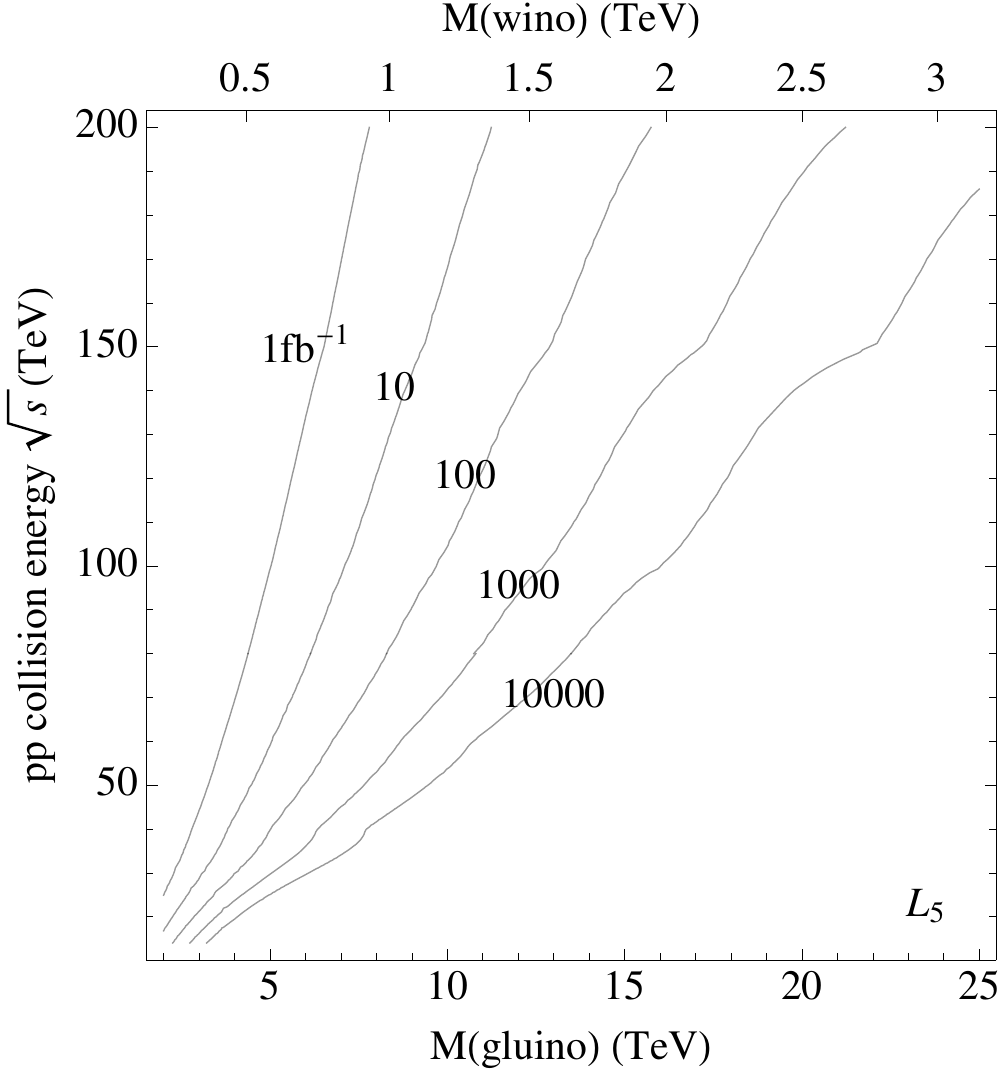} 
}
\caption{Luminosity (fb$^{-1}$) needed to achieve $5\sigma$ statistical significance, ${\cal L}_5$, is contour plotted. The scaling rule in \Eq{eq:scaling} is applied. In the upper horizontal axis, the corresponding wino mass is also shown as in \Fig{fig:gluino-prod}.}
\label{fig:reach-glupair}
\end{figure}

Using the scaling rule and gluino pair production rates shown in \Fig{fig:gluino-prod}, we obtain discovery reach of the AMSB gaugino model as shown in \Fig{fig:reach-glupair}. The luminosity needed for $5\sigma$ statistical significance is contour plotted. The scaling rule is applied to the results optimized for 10 TeV gluino at 100 TeV LHC discussed in previous subsection. We have ignored systematic uncertainties, which are unlikely to change the results by much; NLO K-factor uncertainty and possible branching ratio suppression can have order one effects on production rate, and see appendix for possible simulation uncertainties. Compared to a dedicated recent example study in Ref.\cite{Cohen:2013zla,Bhattacherjee:2012ed}, our significance estimation is a bit more optimistic.

In \Fig{fig:reach-glupair}, we also show corresponding wino mass in the AMSB model with $m_{\widetilde{f}} = m_{3/2}$. To probe all the way up to 3.1 TeV wino LSP, we may need at least 200 \TeV $pp$ collider with ${\cal O}(1000)\, {\rm fb}^{-1}$ of data. If sfermions are lighter and just as heavy as gluinos, about 2-4 times smaller luminosity is needed as the gluino becomes comparatively lighter.

%%%%%%%%%%%
\section{General interpretation}
%%%%%
\vspace{0.1in}
{\bf \emph{Higgsino NLSP.}} 
Our $M_{eff}$ analysis is applicable much more widely, not just to the AMSB model without light higgsinos. If we allow light higgsino NLSP in the AMSB gaugino model, one possible modification is mixing between gauginos and higgsinos. If $|M_{1,2} - \mu| \gtrsim 450\GeV$ with TeV-scale gauginos, the mass splitting between charged and neutral components of wino/higgsino are still kept smaller than 10 GeV making chargino decays inaccessible. In this limit, multi-jet final states from gluino productions dominate most, and this analysis is simplest and most appropriate.

However, pure higgsino NLSP can also enable a two-step decay of the gluinos 
\beq
\widetilde{g} \to \widetilde{h}\, jj \to \widetilde{w}\, jjjj \quad \textrm{where} \quad  \widetilde{h} \to \widetilde{w} \+ h ,\, W, \,Z.
\eeq
If higgsinos are kept far away from winos, more than 85\% of gluino {\it pairs} still decay into pure hadronic final states either via one-step (60-70\%) or two-step decays (20-25\%). In the pure hadronic final states, previous $M_{eff}$ analysis can be used with the same set of backgrounds\footnote{{Higgsino NLSP production is another efficient discovery mode when light higgsinos exist~\cite{ATLAS:2013hta,Han:2013kza}. But, when higgsinos are marginally close to winos, gluinos can be more useful; in addition to $M_{eff}$ analysis (as discussed here), boosted leptons from two-step decays can also be utilized~\cite{Giudice:2010wb,Gori:2013ala}.}}.

Fortunately, two-step decays of the gluino do not modify $M_{eff}$ spectrum significantly compared to that of one-step decays as long as the gluino-to-higgsino mass ratio is greater than about 3. If higgsinos are closer to gluinos, visible particles are softer while missing transverse energy is smaller, and $M_{eff}$ becomes significantly smaller. The difference, however, resides in how much $M_{eff}$ is contributed from visible particles or missing transverse energy (MET). Two-step decay has more visible particles and smaller MET. This can help to distinguish one-step vs.\ two-step decay using the $M_J$ variable discussed earlier, which is larger for messier two-step decay; but this separate measurement will need much more luminosity.

In all, effects of adding NLSP higgsinos to the AMSB gaugino model are minimal as long as higgsinos are far away from gauginos. The small changes in the branching ratios and the $M_{eff}$ spectrum result in about 1.5 times more luminosity needed for the discovery compared to that shown in \Fig{fig:reach-glupair}.

\begin{figure*}
\centerline{
\includegraphics[width=1.9\columnwidth]{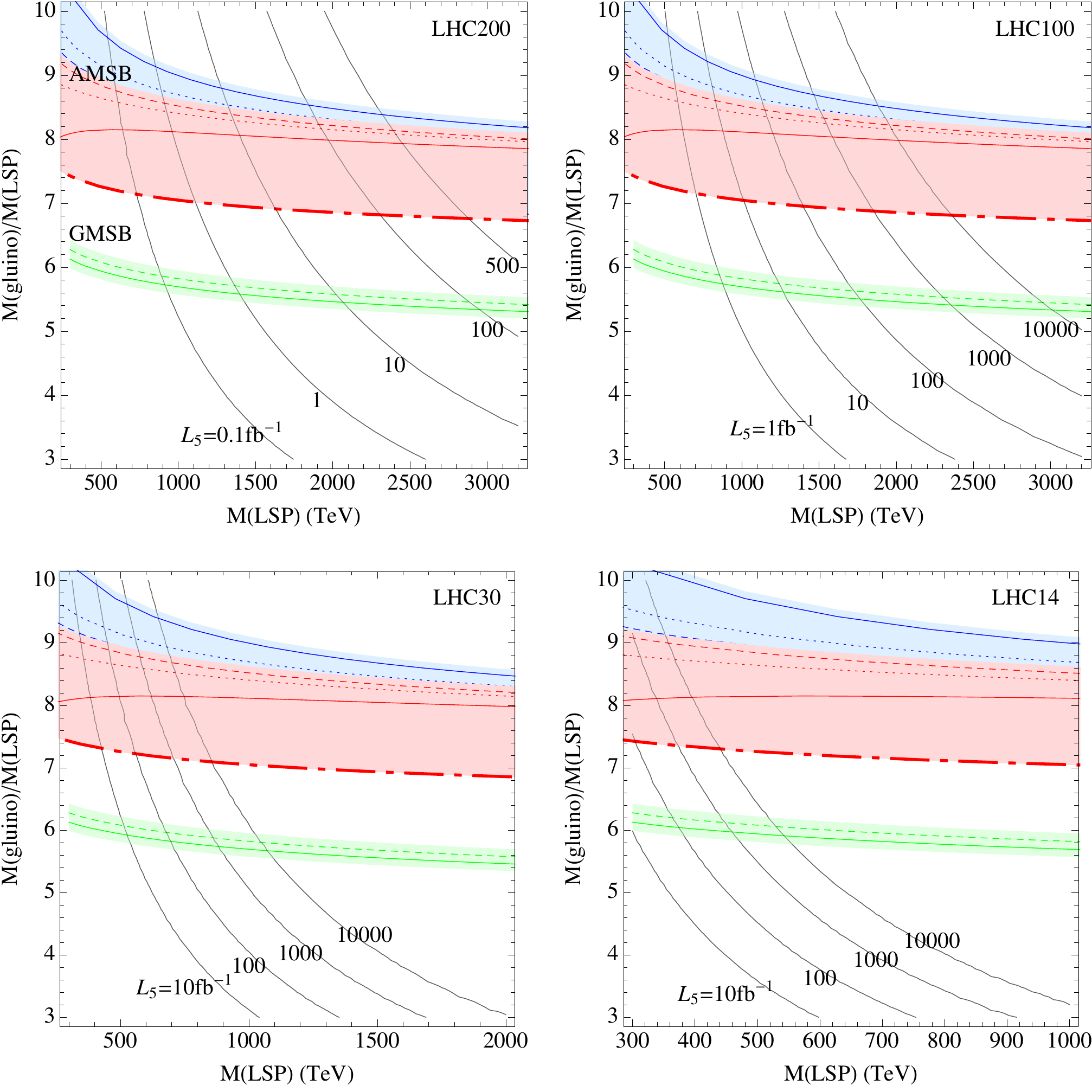}
}
\caption{Contours of luminosity needed for $5\sigma$ statistical discovery, ${\cal L}_5$, in the plane of LSP mass and gluino-to-LSP mass ratio for several $pp$ collision energies 200, 100, 30, and 14 TeV. Blue and red regions are AMSB predictions as in \Fig{fig:gaugino_masses} left panel. mGMSB predictions at NLO are shown in green. Uncertainties of mGMSB are estimated by varying the renormalization scale and the messenger scale ($\Lambda_{SUSY}$(dashed) $\leq M_m \leq M_{Pl}$(solid)). Mirage mediation can in principle take any ratio below the AMSB prediction, but $0.7 \lesssim \alpha \lesssim 8$ predicts gluino-to-LSP mass ratio smaller than 3.}
\label{fig:mratio-reach}
\end{figure*}
%

%%%%%
\vspace{0.1in}
{\bf \emph{Gluino-to-LSP mass ratio.}}
The $M_{eff}$ spectrum of one-step decaying gluinos is not that sensitive to the LSP mass if gluino-to-LSP mass ratio is greater than 3. Again, if a LSP is closer to the gluino, $M_{eff}$ becomes smaller and the signal is buried under backgrounds. Thus, the previous results of $M_{eff}$ are valid for any models of gaugino mass (not just to AMSB) as long as these validity conditions and simplifying assumptions are satisfied; otherwise, $M_{eff}$ would not be the right approach for discovery.

This is very useful because we can interpret our results as the discovery reach of gluino-to-LSP mass ratio in a model independent way, and the gaugino mass ratio is what is sensitive to the underlying SUSY breaking mediation model. {Although different SUSY breaking mediation models may have different golden channels for discovery, we also study others models to illustrate the usefulness of interpreting results in terms of gluino-to-LSP mass ratio.} 

{Three characteristic patterns of gaugino mass ratios are categorized in Ref.\cite{Choi:2007ka} as} minimal supergravity (mSUGRA), AMSB and mirage mediation patterns. In this classification, the minimal gauge mediation model (mGMSB) is subsumed under mSUGRA. We consider mGMSB where scalars are relatively light $m_{\widetilde{f}} \sim m_{\widetilde{g}}$, and calculate gaugino masses. LO gaugino masses are~\cite{Giudice:1998bp}
\beq
M_i^G \= \frac{\alpha_i}{4\pi} \Lambda_{SUSY},
\label{eq:gmsb}\eeq
where $\Lambda_{SUSY} = F/M_m$ in terms of the messenger scale $M_m$ and effective SUSY breaking $F$ term. Here, gauge couplings are GUT-normalized so that $\alpha_1 = \frac{5}{3} \alpha_Y$, and $g_{GUT}^2\simeq 0.5$. {Gauginos are unified at the unification scale.}
The mirage mediation pattern is represented as~\cite{Choi:2005uz}
\bea
M_i^M &=& \frac{\alpha_i}{4\pi} \left( \frac{1}{\alpha} \frac{16\pi^2}{ g_{GUT}^2 \ln (M_{Pl}/m_{3/2})} \+ b_i \right) m_{3/2} \nonumber\\
&\simeq& \frac{\alpha_i}{4\pi} \left( \frac{1}{ 0.1\alpha} \+ b_i \right) m_{3/2}
\label{eq:mirage}\eea
where $\alpha \sim {\cal O}(1)$ is a free parameter {and the factor $1/0.1$ arises from KKLT type models} -- the notation conforms with that of Ref.\cite{Choi:2005uz}. Mirage pattern is effectively an interpolation between the mSUGRA class ($\alpha \to 0$) with $\Lambda_{SUSY}=m_{3/2}/(0.1 \alpha)$ and AMSB ($\alpha \to \infty$) patterns, and gauginos are unified at an intermediate value $\alpha \simeq 2$ {or at some intermediate energy scale}. Thus, mirage mediation models, typically having $\alpha \sim 1$, predict the smallest mass ratios among these models; $0.7 \lesssim \alpha \lesssim 8$  predicts gluino-to-LSP mass ratio smaller than 3, and our discovery reach will not apply well. {But if the mirage pattern is induced really by combination of GMSB and AMSB as in deflected AMSB models~\cite{Pomarol:1999ie,Rattazzi:1999qg} (instead of KKLT type), for example, a wider range of $\alpha$ may be viable.}

We use NLO results in this paper, and refer to Appendix for details of our NLO calculations.

Most patterns satisfy our validity conditions and simplifying assumptions -- exceptions are mirage mediation with $\alpha \sim 2$ where a compressed spectrum is obtained. In \Fig{fig:mratio-reach}, we interpret results of \Fig{fig:reach-glupair} as luminostiy needed for $5\sigma$ statistical significance in the gluino-to-LSP mass ratio and LSP mass. We also show AMSB and mGMSB predictions. Mirage pattern in principle can predict any value of mass ratio below AMSB prediction, but typically predicts mass ratios smaller than mGMSB. 

The errors of AMSB are estimated as previously explained. The mGMSB error bands are estimated by varying the renormalization scale and the messenger scale between $F/M_m \leq M_m \leq M_{Pl}$. The lower bound implies that the spurion superfield coupling method of expansion in powers of $F/M_m^2$ becomes invalid. For mGMSB, uncertainties from the renormalization scale and the messenger scale are comparable. Again, the main dependence on LSP mass comes from the scale dependence of gauge couplings. Errors of mGMSB in \Fig{fig:mratio-reach} is smaller than that of AMSB because scalars are lighter -- heavier scalars will modify mGMSB by larger amount as well.

Combined with model dependent cosmological and astrophysical bounds on LSP mass, the \Fig{fig:mratio-reach} can provide useful discovery prospect of SUSY breaking mediation models. 3.1 TeV wino LSP of AMSB model can be probed at LHC200 with ${\cal O}(1000)fb^{-1}$ of data. It should be kept in mind that astrophysical constraints may be ruling out winos that make up the full cold dark matter~\cite{Cohen:2013ama,Fan:2013faa}, including a 3.1 TeV wino with assumed thermal relic abundance; however, those constraints are dependent on somewhat uncertain halo assumptions, and also dependent upon absolute stability of the wino. Higgsino LSP can also be searched. If 1TeV higgsino is the LSP in a general spectrum, 30 TeV LHC can probe up to 6 TeV gluino (equivalently, mass ratio 6) with ${\cal O}(1000)fb^{-1}$. 500 GeV bino LSP in the mGMSB, for example, can be probed with 1000 fb$^{-1}$ at LHC14.

{The discovery reach in the gluino-to-LSP mass ratio is applicable more generally in a model independent way. For example, variants of GMSB where gauginos follow GMSB relation (with proper NLO corrections) while scalars feel extra SUSY breaking to become heavier can also be constrained in the same way as mGMSB. Another example is a scenario with more general non-universal gaugino masses at a boundary scale (or, the unification scale) that have recently been widely discussed to reduce Higgs mass fine-tuning~\cite{Yanagida:2013ah,Kaminska:2013mya,Martin:2013aha}. Although these types of models may not have discussed gaugino mass patterns, it can still be studied with our approach as long as the gluino is much heavier than its decay products.}

Another useful property of $M_{eff}$ spectrum is that the peak location is most sensitive to the gluino mass~\cite{Hinchliffe:1996iu} as long as the gluino is heavier than lighter gauginos and higgsinos by more than a factor 3; see also \Fig{fig:meff}. How the gluino decays and the masses of lighter inos are not so important. Thus, the location of a peak can be used to measure gluino mass. The LSP mass may be measured in precision measurements using leptons in other two-step decays of gluinos as well~\cite{Hinchliffe:1996iu}, or by LSP pair production at a future linear collider~\cite{Baer:2013cma}.

%%%%%%%%%%%%%%%%
\section{Conclusion}

In conclusion, we have probed the search capabilities of $pp$ hadron colliders for gluinos in the context of split supersymmetry, gauge mediation and mirage mediation. We have found that when the gluino to LSP mass ratio is not more than 10 -- which is expected since gaugino mass ratios are related to each other through their respective gauge couplings or $\beta$-functions in most scenarios -- one can cover the full range of parameters at a 200 TeV collider with approximately 1000 $\xfb^{-1}$ of integrated luminosity. A 100 TeV collider is not enough energy to fully cover the parameter space. Indeed, perhaps the most interesting case of wino LSP thermal dark matter with gauginos satisfying the AMSB mass hierarchy, a 100 TeV collider with even $10,000\,\xfb^{-1}$ of integrated luminosity is not enough.

Our analysis has been based on the $M_{eff}$ approach~\cite{Hinchliffe:2001bz}, with no special kinematic variables for the signal applied, nor did we include third-family tagging analysis, which could be useful if third-family particles are more prevalent in the decays of the gluinos in these scenarios, as might be expected~\cite{Toharia:2005gm,Gambino:2005eh,Acharya:2009gb}.

On the other hand, there is a vast region of parameter space yet to be explored by the 14 TeV LHC and by incrementally higher energy colliders that are still untouched by current experiments, and can have dark matter despite not having the standard thermal history~\cite{Gherghetta:1999sw,Moroi:1999zb}. Therefore, discovery is possible at any time, but definitive coverage of the wino LSP scenario through gluino production and decay requires a $\sim 200\TeV$ collider.

Definitive coverage of the higgsino LSP scenario is not possible at any collider, since supersymmetric gauge coupling unification and thermal dark matter are viable for degenerate higgsinos at $\sim 1\TeV$ with all other superpartners arbitrarily heavy. The gluino mass has less reason to be connected a priori to the Higgsino mass than it had with the wino mass.  Nevertheless, we consider this to be somewhat less likely option since the higgsino mass alternatively would be correlated with the much heavier scalar sector and therefore irrelevant to dark matter concerns and TeV physics, leaving gauginos to play that role. Thus, if it is light, it is likely to be accidentally light and correlated with the gaugino masses, leading to gluino decays to higgsinos as a viable search option. In this case all of our results apply, and ratios of gluino to higgsino can be probed much higher than gluino to wino, only because thermal dark matter higgsinos are much lighter than thermal dark matter winos.

Finally, we remark that our analysis is model independent in as much as gluino masses are much heavier than their primary decay products. Given that gaugino masses are often correlated with the gauge coupling strengths and $\beta$-function values of their corresponding gauge groups, which is the discussed cases of mSUGRA, mGMSB, AMSB and mirage mediation, this strikes us a generic prospect.

%%%%%%%%%%
\vspace{0.1in}
{\bf \emph{Acknowledgement.}} We appreciate K. Jung Bae, K. Choi, T. Cohen, S. Hui Im, P. Ko, H. Min Lee, A. Pierce, Y.-W. Yoon for valuable discussions. S.J. thanks KIAS Center for Advanced Computation for providing computing resources. S.J. is supported in part by National Research Foundation (NRF) of Korea under grant 2013R1A1A2058449.

%%%%%
\vspace{0.1in}
{\bf \emph{Appendix - NLO gaugino masses.}}
LO gaugino masses were presented for the AMSB in \Eq{eq:amsb}, the GMSB in \Eq{eq:gmsb}, and the mirage mediation in \Eq{eq:mirage}.

We include NLO corrections to these relations in this paper. Three types of NLO corrections are consistently added. First, NLO corrections to matching condition at the messenger scale are added for the GMSB. This matching correciton does not exist for AMSB, and the relation \Eq{eq:amsb} is exact to all orders~\cite{Giudice:1998xp}. Including NLO matching corrections, GMSB masses at the messenger scale $M_m$ is given by~\cite{ArkaniHamed:1998kj}
\beq
M_i^G (M_m) = \frac{\alpha_i(M_m)}{4\pi} \left( 1+ T_{G_i} \frac{\alpha_i(M_m)}{2\pi} \right) \frac{F}{M_m}
\eeq
where $T_{G_i} = i$ for $SU(i)$ and $T_{G_i}=0$ for $U(1)$. Gaugino screening theorem is manifest as NLO correction is proportional to gauge coupling itself but not to others. 

A second type of NLO corrections is to use two-loop renormalization group equations of MSSM; we refer to RG equations in Ref.\cite{Martin:1993zk}. Two-loop running resums next-to-leading logs which are formally the same order as finite one-loop corrections. For AMSB, two-loop parts of $b_i$ in \Eq{eq:amsb} give dominant corrections to wino and bino masses~\cite{Gherghetta:1999sw}.

It is these finite one-loop terms that are finally added to complete NLO corrections. This correction defines gaugino pole masses in terms of running gaugino masses. Pole masses are physical observables, and the renormalization scale dependence of gaugino self-energy corrections are cancelled by that of gauge couplings (at one-loop level in our case). In this work, we add threshold corrections explicitly by straighforward one-loop Feynman diagram calculations without constructing effective theories. Our calculations are based on Ref.\cite{Pierce:1996zz}, and agree with that of Ref.\cite{Gupta:2012gu}. We further generalize in case of different orderings of gauginos and higgsinos -- different orderings have different arguments of threshold log terms. 

Mirage gaugino masses are conveniently first matched at the messenger scale of  GMSB (or at the Planck scale if mSUGRA), RG-evolved down, and added with threshold corrections to define pole masses.

When sfermions are very heavy, there appear large log terms so that a proper low-energy effective theory should be constructed by matching and running {as done in Ref.\cite{Giudice:2004tc}}. However, we are content with observing that this uncertainty, by virtue of not specifying the scalar masses precisely, is not large enough to substantively change our results based on the leading order predictions.

%%%%%
\vspace{0.1in}
{\bf \emph{Appendix: Event generation.}}
Generating tails of backgrounds reliably is crucial in this work. For example at 200 TeV collision, 15 TeV gluino pairs give effective mass around 20 TeV, but only $1/10^7$ fraction of $W+j$ backgrounds have this high effective mass. 

Similarly to Ref.~\cite{Hook:2012fd,Hedri:2013pvl,Avetisyan:2013onh}, we divide phase space with successively smaller cross-section, and generate similar number of events in each sectors. We find it useful to use scalar sum of final state jets to divide phase space in our study. In decreasing order of cross-section, we denote each of the sectors of phase space by $P_i$. The choice of cross-section ratio $\sigma(P_{i+1})/\sigma(P_i) = 0.9$ seems resonable and convenient. We generate $W/Z/t\bar{t}+0,1,2j$ backgrounds using {\sc{MadGraph}} for each of following sectors divided in terms of $H_T(j)$ according to the aforementioned cross-section ratio (at $\sqrt{s}=100$TeV):
\bea
W+j : \, H_T(j) \= \{ \, 0, \cdots,  700, 1500, 2800, 5200, \nonumber\\
8800, 13800, 20500\GeV  \} 
\eea
\bea
Z+j : \, H_T(j) \= \{ \, 0, \cdots,  650, 1400, 2800, 5000, \nonumber\\
8500, 13500\GeV  \} 
\eea
\bea
t\bar{ t}+j : \, H_T(j) \= \{ \, 0, \cdots,  750, 1500, 2600, 4300, \nonumber\\
6700, 10000, 14500\GeV  \} 
\eea
We omit to list phase space sectors that are not relevant to multi-TeV ino searches in our study. Other generation-level cuts are jets : $p_T(j)>50,\, \eta<5$ and minimal separation between jets and leptons. These cuts improve computation efficiency. On the other hand, signal samples are generated all at once without generation-level cuts. 

Up to two additional partons are generated with {\sc MadGraph} and matched with {\sc Pythia} parton showered results using MLM matching scheme~\cite{Mangano:2006rw}. A merging scale $\mathbf{xqcut}=50$GeV is used~\cite{Alwall:2007fs} for all samples. No hadronization, nor multiple interactions are simulated. {A recent dedicated study including these effects in Ref.~\cite{Cohen:2013zla} shows small difference in discovery estimation -- ours are somewhat more optimistic though.} Our 40 TeV background estimation (generated in the same way) agrees well with the experimental study reported in Ref.~\cite{Hinchliffe:2001bz}.

To define jets and isolated leptons in the messy environment of high-energy $pp$ collision, we first jet-cluster all energy deposits using {\sc FastJet}\cite{Cacciari:2011ma} anti-$k_T$\cite{Cacciari:2008gp} with $R=0.4$ producing a list of proto-jets. If a proto-jet contains a lepton whose $p_T$  is higher than 50\% of the proto-jet's $p_T$, we assign the proto-jet to be an isolated lepton. All remaining proto-jets are assigned to be normal hadronic jets. No detector simulation. Fat jet simulation used to define $M_J$ is carried out with anti-$k_T$ $R=1.0$.

%%%%%%%%%%%%%%%%%%

\end{document}